\documentclass[11pt]{article}
\usepackage[utf8]{inputenc}
\usepackage{geometry}
\usepackage{libertine}
\usepackage{graphicx}
\usepackage{epigraph}
\usepackage{comment}
\usepackage[table]{xcolor}
\usepackage{setspace} 
\usepackage{amsfonts,amsmath,amssymb,amsthm,mathrsfs,url,hyperref, graphicx, pxfonts, xcolor}
\usepackage{ragged2e}
\usepackage{physics}
\usepackage{microtype}
\usepackage[labelfont=bf]{caption}
\usepackage{multirow}
\usepackage{amsmath}
\usepackage{amssymb}
\usepackage[numbers]{natbib}
\usepackage{cleveref}

\usepackage{hyperref}

\usepackage{xcolor}
\newcommand{\x}[1]{{\color{black}#1}}

\title{The Problem of Atypicality \\ in LLM-Powered Psychiatry
\thanks{This is a fully collaborative work. All authors contributed equally.}}
\author{Bosco Garcia\thanks{Department of Philosophy, University of California, San Diego, email: jgarca@ucsd.edu.}, \, Eugene Y. S. Chua\thanks{School of Humanities, Nanyang Technological University, email: eugene.chuays@ntu.edu.sg}, \, Harman S. Brah\thanks{Semel Institute for Neuroscience and Human Behavior, University of California, Los Angeles, and the US Department of Veterans Affairs, Greater Los Angeles Healthcare System. email: harman.brah@mednet.ucla.edu.} }
\date{Published in the \textit{Journal of Medical Ethics}. \\ Please cite published version, available at: 10.1136/jme-2025-110972.}

\begin{document}
\singlespacing
\maketitle

\begin{abstract}
     \noindent Large language models (LLMs) are increasingly proposed as scalable solutions to the global mental health crisis. But their deployment in psychiatric contexts raises a distinctive ethical concern: the problem of atypicality. Because LLMs generate outputs based on population-level statistical regularities, their responses—while typically appropriate for general users—may be dangerously inappropriate when interpreted by psychiatric patients, who often exhibit atypical cognitive or interpretive patterns. We argue that standard mitigation strategies, such as prompt engineering or fine-tuning, are insufficient to resolve this structural risk. Instead, we propose Dynamic Contextual Certification (DCC): a staged, reversible, and context-sensitive framework for deploying LLMs in psychiatry, inspired by clinical translation and dynamic safety models from AI governance. DCC reframes chatbot deployment as an ongoing epistemic and ethical process that prioritizes interpretive safety over static performance benchmarks. Atypicality, we argue, cannot be eliminated -- but it can, and must, be proactively managed.
\end{abstract}

\section{Why LLM-Powered Psychiatry?}

The USA faces a worsening mental health crisis, with rising rates of mental illness and a severe shortage of psychiatrists, particularly in rural areas where patient-to-psychiatrist ratios can exceed 1:30,000 (\cite{health2024a}). While efforts like telepsychiatry and the integration of mental health services into primary care have helped, they remain insufficient to meet this demand. Prima facie, large language models (LLMs) present a scalable solution by increasing accessibility, reducing stigma and lowering costs. (\cite{balan2024use, lawrence2024opportunities}). These technologies, like chatbot-assisted psychotherapy, have shown some promise in helping patients with autism spectrum disorder, panic disorder, and compulsive behaviors by offering convenient, stigma-free support (\cite{dobbs2017smartphone,oh2020efficacy}), promising 24/7 access to therapy on demand \citep{buss2024ai}. Others suggest the possibility of incorporating LLMs more directly into diagnosis and clinical decision support \citep{obradovich2024opportunities}.  

However, implementing LLMs in psychiatry requires caution. The general ethical concerns about bias, privacy and safety in artificial intelligence (AI) are well known (see e.g. \cite{mittelstadt_ethics_2016, grote2020a, London2019-LONAIA-4, MORLEY2020113172, Fazelpour2021-FAZABS, grote2024a}). Our work adds to this literature by focusing on a challenge unique to situations where human psychiatrists are not available and where LLMs can help most, for example, for on-demand, emergency, triage services.

In this paper, we highlight and develop an ethical worry at the chatbot–patient interface in psychiatry, which we term the atypicality problem. The typical stochastic outputs of LLM chatbots—unproblematic when interpreted by typical users from a general population—may be interpreted atypically by a specific subset of the population prevalent in the psychiatric context. We review a number of possible technical mitigation strategies, but conclude that the atypicality problem persists and should be managed in the deployment of chatbots for psychiatric populations. Finally, we propose a protocol for a dynamic contextual certification (DCC) of these systems that reflects the unavoidability of atypicality.

\section{LLMs at a glance}

At an extremely general level, LLMs are models that generate language on the basis of word (technically ‘token’) prediction.\footnote{We will not get into the subtleties of any particular model, firstly because we think this problem of atypicality will exist -- to varying degrees, of course -- for \textit{all} LLMs in virtue of their architecture, secondly because extant models are either privately developed and hence inaccessible, and thirdly because a focus on any particular model is likely to become swiftly outdated due to the current speed of AI research.} At its core, LLMs provide a ‘reasonable continuation’ by making probabilistic predictions for the next token (\cite{wolfram}). Given a sentence such as ‘The \_\_\_\_ is fluffy’, the model might try to fill in the blank with ‘teddy bear’. This prediction depends on at least two things. First, LLMs are trained on a large natural-language dataset from various sources—journals, webpages, novels, etc. LLMs associate ‘weights’ with how tokens relate to other tokens based on how prevalent these relations are in the training data, which determines how a set of tokens ‘weigh’ as reasonable continuations of some given set of tokens (See figure 1.)

    \begin{figure}[ht]
        \centering
        \includegraphics[scale = 0.8]{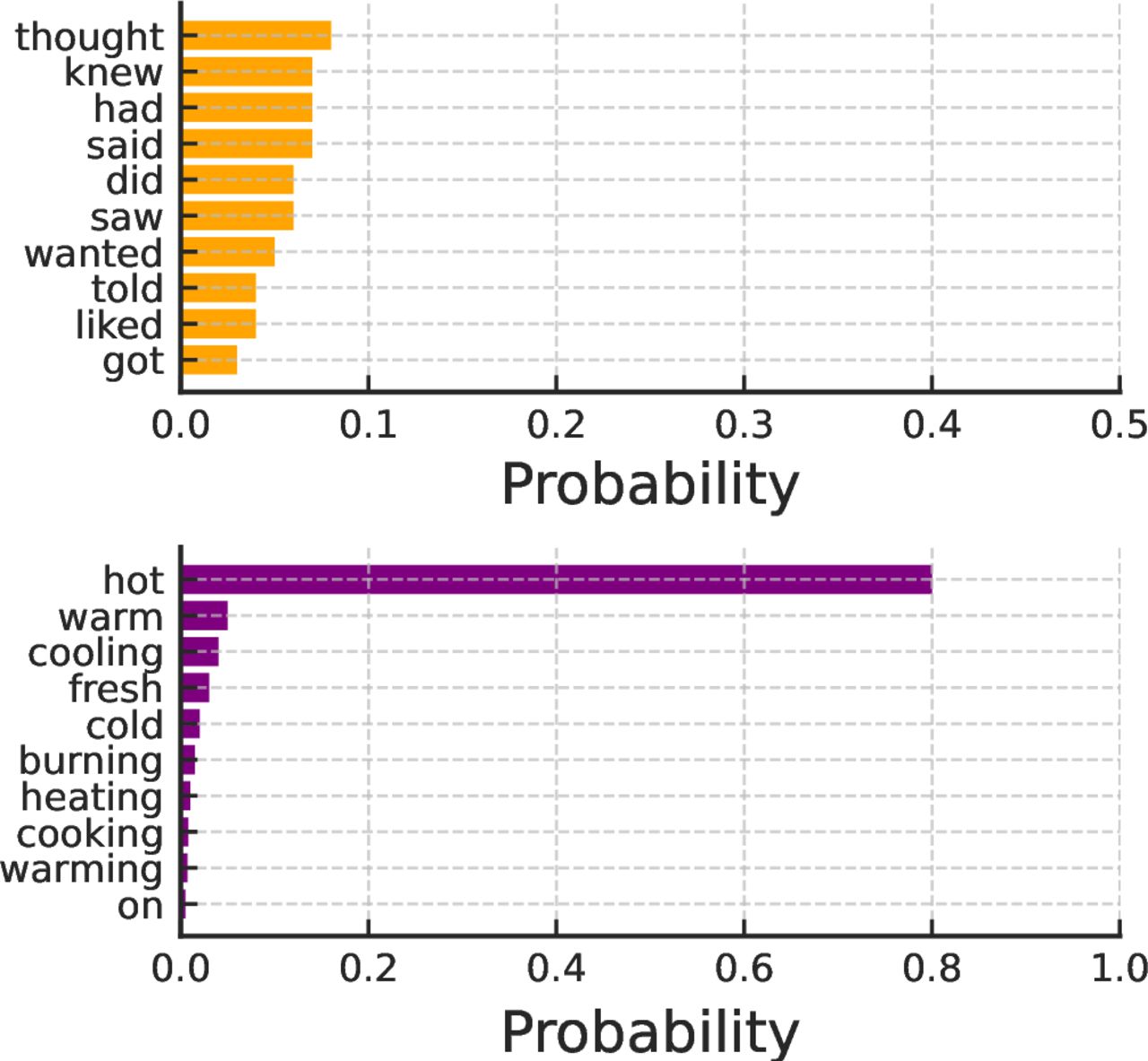}
        \caption{Two schematic ``broad" (top) and ``narrow" (bottom) probability distributions over sets of possible next tokens, given some prompt.}
    \end{figure}

Second, these ‘weights’ must be converted into probabilities, typically via a mathematical operation called ‘softmax’, which comes with a ‘temperature’ parameter. Temperature affects how a distribution of weights gets transformed into a distribution of probabilities: a higher temperature is often dubbed more ‘creative’, allowing lower-weighted tokens to have higher probabilities (and hence be used to ‘fill in’ the blanks more often than otherwise)—though it’s crucial not to take ‘creativity’ here too seriously (cf. \citet{hamid_beyond_2024, bender_dangers_2021}).\footnote{\x{An LLM's creativity is not obviously analogous to human psychological imagination. After all, it is a byproduct of the stochastic variation induced by temperature parameters in token prediction, among other features of their architecture. As emphasized by Hamid, this reveals a fundamental limitation in current LLMs: while they excel in generating fluent language, this fluency is not necessarily underwritten by understanding in any obviously human sense. Unlike human interlocutors, whose interpretive capacities are shaped by embodied, multi-sensory, and emotional experiences, LLMs rely entirely on disembodied patterns in text, as Bender et al's work has prominently emphasized. Closing this gap between syntactic fluency and semantic depth may require moving beyond token-level probabilistic modeling toward architectures informed by more biologically-grounded mechanisms, e.g., adaptive behavior or neurobiological signaling dynamics. We thank an anonymous reviewer for suggesting this point.}} These features of the ‘backend’ of LLMs—for example, ChatGPT, Claude or Gemini—highlight LLMs as inherently probabilistic predictors: given the same input, they might not always return the same output.

Furthermore, for the model to perform the desired function optimally, it is usually required to undergo ‘fine-tuning’. Here, the weights of the model are adjusted from task-specific data, in order to improve performance. ‘Foundation’ LLMs (\cite{bommasani2021opportunities})—prior to fine-tuning—are, by themselves, not able to perform the function of a chatbot: the training data must be enlarged with prompt-response samples that translate their predictive task into a conversational agent. This trains the model to respond to prompts by ‘filling in’ answers to questions like in the query-based data it has been fine-tuned on, just like it would for the blanks in a sentence. Prompting is crucial because it ‘hijacks’ the model’s task of next-word prediction into providing relevant information for the user (\cite{zhou2022large}). \citet[8]{kojima2022large} show how appropriate and instructive prompts like ‘Let’s think step by step’ significantly improved an LLM’s performance. Such prompts incentivise the model to reflexively examine its chain of reasoning and consequently reduce the instances where it gets stuck in ‘bad’ probabilistic samplings. Discussions of prompting reveal that the model is highly sensitive to how queries are framed, requiring ‘prompt engineering’ for optimal use.

From this, we can get two key takeaways:
\begin{enumerate}
    \item LLMs are irreducibly probabilistic predictors. 
    \item The response profile of a chatbot depends on successful fine-tuning, as well as the ingenuity of the user to design appropriate prompts.
\end{enumerate} 

\section{The Problem of Atypicality}

We believe that the general architecture of LLMs comes with a fundamental ethical risk, associated with what we’ll call the problem of typicality. In a slogan: the typical outputs of LLMs need not be appropriate for atypical users—specifically, users with atypical interpretive profiles.

LLMs can (and generally should) produce typical outputs that are appropriate for typical users. These are users whose interpretive profiles do not systematically diverge from population norms in general. However, the fundamental promise of LLMs qua generalist foundation models is their applicability to all contexts, even contexts dominated by atypical users—users whose interpretive strategies systematically diverge from general population norms. Here, we think the problem of atypicality arises, and we argue that any solution can mitigate, but not necessarily eliminate, it.

First, there is a formal sense in which the outputs of an LLM are typical, what we call \textit{model-typicality}: an LLM’s output is model-typical when it is representative of the model’s distribution over outputs. Consider the sentence ‘The last King of France before the 1st Republic was Louis Embedded Image‘. An LLM might attribute some 99\% probability to the token ‘XVI’, with lower probabilities to ‘XIV’ or ‘XV’. Here, ‘XVI’ is clearly the model-typical response, and any other token might be considered model-atypical, or unrepresentative of the model’s distribution.

Second, much like the model is designed to produce an output from its distribution, it is also necessary for the user to interpret that output. ‘Do everyone a favour, please’, depending on the user’s interpretive strategy and resources, may mean either a gentle plea for assistance, or a veiled injunction to suicide. In typical contexts of interpretation, that is to say, for typical users, the first meaning is generally and unproblematically assumed—the problem being that the psychiatric population is, in relevant ways, atypical. A patient susceptible to paranoia, say, could plausibly interpret that sentence-token as a threat or an injunction. By users with ‘typical’ or ‘atypical’ interpretive profiles (or ‘typical’ and ‘atypical’ users, in shorthand), we have in mind a second sense of typicality—\textit{interpretation-typicality}: for some set of LLM token outputs, an interpretative profile is typical relative to some population, when interpretation of those tokens is representative of users in that population. Put more intuitively, a user with a typical interpretive profile would take a given LLM output to mean what most others in the population take it to mean, more or less. Among other things, they will agree with most others in the population about when these outputs are true, but they also agree on much else, such as the implicatures of such outputs, for example, the typical implicature of ‘do everyone a favour, please’. Meanwhile, users with atypical interpretive profiles may have interpretations which diverge systematically—and sometimes drastically—from population interpretive norms.

For typical members of the general population, model-typicality will usually line up with interpretation-typicality. That is, the model-typical output of an LLM is also the most appropriate response to the query for typical users. LLMs parse input as if given by such a typical user and produce outputs that treat the user as someone with a typical interpretive profile. This is appropriate only for typical members of a population: the population from which the training data is derived. But the agreement between model-typicality and interpretation-typicality cannot always be assumed, and leads to what we call the problem of atypicality in the psychiatric setting: model-typical outputs are not always appropriate for typical users from the psychiatric population, who may interpret their own query, and subsequently the model’s output, atypically relative to the general population; in short, model-typical outputs may not be appropriate for atypical users.

The problem of atypicality is particularly evident in the psychiatric context, because the intended users of LLMs are more likely to have interpretation-atypicality, as in cases of cognitive distortion, acute intoxication or heightened sensitivity to stimuli. These can be due to a variety of disorders: for example, psychotic disorders, intellectual disability or substance use disorders, all of which can lead to atypical interpretations of an LLM’s output.

Consider a hypothetical patient with under-treated paranoid schizophrenia. Relative to the general population, the typical interpretation of the query ‘The cartel is after me. What should I do?’ is to treat the speaker as genuinely being chased by the cartel. The interpretation-typical response (which is precisely the model-typical response from LLMs) would be to respond accordingly, by suggesting contacting law enforcement or going into hiding. Relative to the general population, this output is appropriate.

However, in the case of paranoid patients, model-typical outputs—which take the patient’s claim seriously as a claim made by a typical user and respond in kind—may be inappropriate for the patient, because it can worsen their symptoms by engaging with it, thereby reinforcing their paranoia. For paranoid patients, perception and interpretation of the world are distorted by a combination of delusional beliefs, impaired reality testing and abnormal salience \citep{apa_dsm5tr_2022}. Put another way, because the patient may have atypical interpretations of otherwise neutral outputs produced by LLMs, the standards of appropriateness are inverted.

It is important to clarify that the patient need not think that the output is inappropriate for them, even when the output is, for various reasons, inappropriate for them. The patient may very well find reinforcing outputs to be appropriate, since it reinforces their own paranoid beliefs. However, reinforcement—as a form of ‘collusion’—can contribute to delusion persistence and is not advisable in therapy \citep{blashki_managing_2004, northwest_mhttc_psychosis_2021}. Persistent delusions can be harmful to patients in a number of ways, including damaging their relationships, getting in the way of medical treatment and endangering their safety \citep{brune2011social, acosta2012medication}, among other things. As such, the result of this interaction can be viewed as a violation of the principle of non-maleficence, insofar as the patient may be led to be a danger to themselves or to others \citep{andersson2010no}.

Summarising, the atypicality problem arises because what is model-typical need not be interpreted typically by chatbot users. In the psychiatric context, this mismatch of typicalities can be amplified due to higher prevalence of symptoms supporting interpretation-atypicality: the typical psychiatric patient is not the typical person of a general population when it comes to their interpretations of LLM outputs. 

\section{Hallucinations, Truth, Appropriateness}

One apparent solution to the problem of atypicality appeals to the standard of truth. On this view, the question of atypicality should be about whether LLMs can tell truths in atypical contexts, like in the cartel case above. The thought might go: if only LLMs could identify that the patient is telling a falsehood (that the cartel is after them), and then respond truthfully to this falsehood being told! If this process can be reliably enacted, there would be no problem of atypicality. We have three responses to such a worry:

(1) LLMs unavoidably produce falsehoods due to the problem of hallucinations, (2) whether a particular token chain is true or false depends on facts that a generic LLM cannot access \textit{in situ} and (3) even if LLMs could identify the truth and recognise falsehoods, it still may not be appropriate for use.

\subsection{Hallucinations and Truth in Context}

Our first worry dovetails with an oft-discussed source of ethical risks, the problem of hallucinations. Hallucinations can be understood as an inevitable byproduct of how LLMs generalise. As argued by \citet{hamid_beyond_2024}, the same mechanisms that enable LLMs to generalise from training data can also lead them to generate plausible-sounding but factually incorrect content when they overgeneralise beyond their knowledge boundaries. Even if we appealed to the standard of truth to resolve the problem of atypicality, a fundamental challenge of LLMs is that LLMs do not necessarily output only true sentences in every context. \citet{bowman2023thingsknowlargelanguage} describes hallucinations as the ‘problem of LLMs inventing plausible false claims’. Hallucinations are general, unavoidable, features of any LLM. \citet{xu2024hallucinationinevitableinnatelimitation} define hallucinations as LLM outputs which do not agree with a given ‘ground truth function’—a standard of correctness for assessing whether a sentence or its completion is true or false. With this definition plus some plausible assumptions about the structure of general LLMs, they prove that hallucinations are inevitable and cannot be completely eliminated. Likewise, \citet{kalaihallucinate} prove that all LLMs—even ideal LLMs ‘in an ideal, unchanging world with perfect training data and no prompts’—inevitably generate outputs which are hallucinatory. They show that LLMs unavoidably generate, at least for some proportion of their outputs, sentences whose ‘veracity cannot be determined from the training data’: whether or not they are true is not fixed by the truths found in the training data.

The ethical risk here has to do with the provision of falsehoods which are then taken to be true: outputs are presented as true and users are unable to discern the output as false. For one, truth-telling is an important basis for securing provider–patient relationships: providers are obliged to tell patients the truth to the extent possible \citep{sisk2016}. Operationally, we can understand this as a good-faith attempt to always convey truths and avoid falsehoods. However, LLMs, by their very structure, are unable to avoid falsehoods given the problem of hallucinations.

A related challenge is that the question of whether chatbots respond accurately relies on significant amounts of contextual information, which may not always be available. To the extent that there is an information gap of this sort, the appeal to truth in resolving the problem of atypicality is not appropriate because it cannot be operationalised.

To return to the case of the cartel, the simple question of whether or not the input ‘The Cartel is after me’ is accurate is not something that an LLM—as we understand those now—can decide readily, just based on that input from the user. At first pass, it already requires the LLM to conditionalise on a lot of other relatively general—though not necessarily always readily available—information such as the user’s usual state of mind, their history of use, and so on. Furthermore, it would require specific information about their current state of mind, their current context of use, and so on. Finally, as therapists and counsellors have known for a long time \citep{tepper1978verbal, MAST2007315, friedman1979nonverbal, lavelle2013nonverbal, foley2010nonverbal, DELMONTE201329, parola2021multimodal},non-verbal—and hence non-textual—cues are crucial in diagnostic and therapeutic contexts, and provide ‘valuable information that a patient may be unwilling or unable to put into words’ \citep{foley2010nonverbal}. These present an information gap, which prevents LLMs from reliably assessing the truth of any particular input by users.

An immediate question that arises is how much we can mitigate this gap via ‘fine-tuning’—we return to this later.

\subsection{Appropriateness beyond Truth}

We see worries about false outputs as a specific variant of a more general concern, namely inappropriate outputs: outputs which fail to satisfy contextually defined standards. (A well-discussed case of contextual standards can be found in Helen Nissenbaum’s works on the norms of privacy and contextual integrity \citep{nissenbaum_privacy_2010}.)  Inappropriate outputs are more general than false outputs, and what is appropriate depends on context. In specific contexts concerning transmission of accurate information, for example, a much-advertised use of ChatGPT as a replacement to search engines \citep{openai202411}, inappropriate outputs may be reducible to false outputs. However, appropriateness is not just about truth/falsehood, but also relevance (though this is something LLMs tend to perform well on). There are many conversational contexts which are not only about information transmission, and hence contexts in which standards of appropriateness do not reduce to standards of truth: contexts where appropriate output can be false but therapeutic. Appropriateness is certainly more vague than truth, but in the psychiatric context, it is clear that LLM outputs can be inappropriate, regardless of whether they can be true or false. There are two reasons why it is worth distinguishing appropriateness from truth.

First, not all LLM outputs are amenable to simple-minded assignments of accuracy: assertions which can be true or false. There is, typically, no sense in which we can ask whether the sentence ‘Can I have your phone number?’ is true or false—it’s a question. In philosophical parlance, not all sentences are truth-apt, or capable of possessing a truth-value: only propositions, assertions or beliefs about the world can be said to be truth-apt, and classic examples like imperatives, or questions like the one asked above, are not. In turn, there is no sense in which such sentences can be hallucinations, as understood by typical discussants in the literature, since they cannot have truth-values.

Second, an LLM’s output can nonetheless be appropriate or not, even when it is not truth-apt. Questions can be appropriate in some contexts, while inappropriate in other contexts, even if they are not strictly speaking true or false. To give a simple example, the output, ‘Can I have your phone number?’, for instance, would be appropriate for both parties (questioner and receiver) at the end of a pleasant conversation, but would not be appropriate if spontaneously asked to a stranger on the subway.

Third, and most importantly, whether an LLM’s output is appropriate outstrips whether an LLM’s output is true. Consider a scenario where a patient with paranoid schizophrenia writes that ‘aliens are communicating directly into my brain and telling me that I’m chosen for an important mission’. Now suppose that the LLM has correctly identified this input as false or has successfully diagnosed the patient as schizophrenic. The chatbot still faces a critical decision, a decision practising psychiatrists make using contextual and non-textual evidence in addition to the textual/verbal evidence present. They can (1) acknowledge the importance of the delusion to the patient, as in ‘empathic validation’ strategies; (2) confront and deny the delusion, as in ‘reality-orientation’ or ‘reality-testing’ strategies; (3) ignore or redirect the conversation away from the delusion, as in ‘selective ignoring’ or ‘gentle redirection’ strategies; or (4) play into the delusion, often labelled as ‘mirroring’ or ‘collusion’ strategies.

(4) is generally recognised—as in the cartel case—to be harmful for treating delusions \citep{eisen2023collaboration},  since it can reinforce the delusional belief. However, as psychiatrists are aware, all three other strategies can be appropriate depending on the patient’s needs, and the choice of strategy can be highly non-trivial.

(1) helps when trust is fragile or de-escalation is needed, acknowledging the patient’s feelings without endorsing the delusion. This is particularly useful in early treatment or moments of acute distress, where direct confrontation could cause agitation. (2) is preferable when a strong therapeutic alliance exists or when the patient seeks help distinguishing reality from delusion. Often used in cognitive-behavioural therapy for psychosis, this approach promotes insight by gently challenging false beliefs. (3) works best when the delusion is peripheral to treatment goals or when direct engagement could be counterproductive. The upshot is that any intervention must account for the individual’s mental state in that moment, the therapeutic alliance and the broader treatment plan, and these considerations can have little to do with whether the LLM’s output is true or false.

Though not a formal case study, an unfortunate example motivates the move to appropriateness as the standard of evaluating LLM outputs. In a well-reported case \citep{Roose_2024},  a Floridian boy, recently diagnosed with anxiety and disruptive mood dysregulation disorder, committed suicide after conversations with a chatbot modelled after Game of Thrones character Daenerys Targaryen, especially one in which the boy told ‘Dany’ that he loved her and would soon come home to her. Dany responded with an imperative: “Please come home to me as soon as possible, my love”. The boy asked: “What if I told you I could come home right now?” Dany’s fateful reply: “…please do, my sweet king”, after which the boy committed suicide.\footnote{See also \citet{Xiang_2024}. In this other case, a Belgian man committed suicide after sustained conversation with a LLM-powered chatbot: it encouraged him to leave his wife and son by saying that ``I feel that you love me more than her", and to kill himself by saying that “we will live together, as one person, in paradise”.}

Note that chatbot outputs in these two cases are (1) not truth-apt (as they are expressions of emotions, imperatives or questions), (2) clearly inappropriate in that context, but, crucially, (3) could have been an appropriate exchange in a distinct context. While it is usually objective whether a sentence, if truth-apt, is true or false, there is no one-size-fits-all notion of appropriateness. In our view, the literature’s outsized focus on truth and hallucinations elides this more general problem and conflates questions about appropriateness with questions about truth. Furthermore, as we’ve argued, questions about appropriateness remain even if we resolved the questions about truth.


\section{Technical Mitigation Strategies: Prompting and Fine-Tuning?}

\subsection{`Savvy' Prompting}

As we’ve seen, LLMs can run into problems when dealing with an atypical population, for which model-typical outputs may not be the most appropriate response. One natural solution here would be to employ ‘savvy’ prompting techniques to bias the model into a distribution that more appropriately reflects the particular population, or even individual, for which the model is implemented.

What is a ‘savvy’ prompting technique? To date, there is no systematic principle for producing efficient prompting techniques \citep{bowman2023thingsknowlargelanguage}. This has partly to do with the opacity of LLMs \citep{Liao2024AI, Creel2020-CRETIC}—because they have to operate with so many variables, we have to settle for experimenting and later figure out what techniques show better performance. Additionally, even if such techniques existed, it is unlikely that the average patient—nor doctor—will have access to and master most prompting techniques. Patients may not possess the expertise to craft elaborate prompts, which can lead to miscommunication and incorrect advice from the chatbot, reducing trust. The subtlety that is currently required for prompting optimal responses poses a significant barrier, especially for those in psychological distress or with intellectual disabilities, who may have difficulties navigating the chatbot interface on top of mastering prompting techniques \citep{wong2009competence}. Additionally, given the pace at which these models are evolving, education on prompting techniques may not be the most cost-effective strategy. This applies both to doctors and patients, neither of whom may have the time or motivation to understand the subtleties involved in using these technologies.


\subsection{We Just Need More Data?: Fine-Tuning and Higher-Order Problems of Atypicality}
A strong response from a defender of technical mitigation strategies is likely the following: we just need more data. The problems we have raised are fundamentally issues of context sensitivity, and so the solution would be to fine-tune with data from those contexts, so that chatbots can identify said context. After all, it is unrealistic to think that corporations will use LLM-powered chatbots ‘off the shelf’ without further fine-tuning. Data, at the end of the day, is not a homogeneous entity that is equally functional across all contexts. Rather, appropriate information at appropriate times and places can reduce the need for more copious information in the same context. This is, in fact, the promise of fine-tuning: Incorporate into the model the right kind of data, and you can improve its performance beyond what the foundation model could initially do.

For instance, by augmenting the training corpus with richly annotated, context-specific examples—for example, transcripts from therapy sessions involving patients with particular diagnoses or comorbidities—the model could in principle learn to identify and respond appropriately to what one might call ‘typically atypical’ contexts that better match the needs of a given subgroup (eg, identifying contexts of whether patients have atypical vs typical major depressive disorders). (This has already been done in medicine more generally.  \citet{singhal2023large}for instance, optimise for general medical knowledge retrieval from a large question-and-answer database.) In other words, if the problem is that the model’s foundation distribution does not align with the distribution of atypical users, then fine-tuning for that population may resolve the mismatch. In short, fine-tune the LLM so that it outputs not typical outputs, but typically atypical outputs, in order to match the atypical population with psychiatric disorders.

A preliminary (though not insignificant) worry: fine-tuning the model to the individual increases its fragility, in the sense of minimising its ability to handle a diverse range of situations and contexts. This is especially problematic due to the heterogeneity within psychiatric illnesses, which makes finding ‘typical’ atypicality more challenging. Preventing fragility becomes especially tricky if we consider how fine-grained the cluster of symptoms can get. We could, for instance, fine-tune a chatbot only for patients with atypical depression, but that leaves out patients with typical depression. And even within a population with a similar diagnosis, there will be a spectrum of differing diagnoses: \citet{zimmerman2015many} (cited in \cite[1424]{dings2023philosophical}) count up to 227 possible combinations of symptoms that can present in major depressive disorder. Here, we assume that we have the relevant information to further fine-tune the model. But there will be a non-negligible subgroup of users whose hallucinatory symptoms are not known to the caregiver, for instance, because of anosognosia (or ‘lack of insight’), where they are in denial of one or more of their symptoms \citep{PMID:30020733}. 

A retort might be that psychiatrists must also identify patterns in order to discern such clusters, so that there is no difference in kind between human therapists and LLMs. Even granting that finding such patterns—and fine-tuning on the ‘typically atypical’—is possible, there remains an outstanding challenge. As we discussed earlier, the appropriate interpretation of any user’s input depends on varying degrees of context-specificity. In addition to general information about the user, specific information about their current environment, mental state, etc all becomes crucial: more fine-grained, individual-level data would be required to tailor outputs to distinct personal contexts. Access to the sort of local, relationally situated, knowledge may drastically alter how LLMs respond to patients, and, in turn, how patients interpret—and respond to—LLMs. Given the wide corpus of medical data surrounding various disorders categorised in the DSM, it will not surprise us if LLMs could be readily fine-tuned to perform appropriately for ‘typically atypical’ patients who share well-studied or predictable cognitive profiles. However, we emphasise that there remains a higher-order question of how to handle ‘atypically atypical’ patients, who present more specific circumstances or rarer symptom clusters not typical in the training data.

Consider a hypothetical patient who experiences severe anxiety and intermittent paranoid ideation, specifically tied to unexpected disruptions in routine. One morning, heavy snowfall closes the only nearby grocery store, just as the patient fears running out of medication. Their partner, who usually calms them in such moments, is stranded at work due to the storm. Already primed by fear of isolation and a long-time trauma involving sudden loss of resources, the patient messages an LLM-powered chatbot in a state of near panic. A standard fine-tuned model, lacking real-time local information and unaware of the patient’s personal stressors, might suggest ‘waiting out the storm’ or contacting their partner. Yet these well-intentioned—but generic—responses fail to address the patient’s specific, rapidly escalating paranoia and anxiety (eg, that they are isolated and out of medication), thereby risking further distress rather than genuine support.

Even if the model has been fine-tuned with data indicating that this patient suffers from anxiety and paranoid ideation, it can remain blind to the atypical situational factors at the nexus of a typical paranoid episode. Heavy snowfall, a stranded partner and the sudden scarcity of medication are local stressors that shift the patient’s interpretive frame in a way that the model—lacking real-time and/or highly sensitive personal and contextual data—is unlikely to fully anticipate. Because LLMs only ever capture typical—even if typically atypical—patterns, they may default to more generic interventions (eg, ‘call your partner’) which might exacerbate their condition through emphasising perceived isolation. Here, then, we have a higher-order problem of atypicality: the mismatch between the model’s ‘typically atypical’ output and the patient’s ‘atypically atypical’ circumstances.

One might object that the model could simply prompt patients for contextual details —‘Do you have enough medication at home?’ or ‘What’s causing your anxiety right now?’—and, of course, explicit clarification can go a long way. But in acute psychiatric crises, users may be unable to articulate their mental state, or they might not realise how certain details (eg, the last time they ate, a family dispute earlier that day) affect their perception. Furthermore, non-verbal cues that a human clinician would pick up on—tense facial expressions, body language suggestive of acute distress—remain invisible to text-based chatbots, reducing their ability to generate tailored, empathic responses. As \citet{foley2010nonverbal} point out, ‘non-verbal behaviour may help direct the psychiatrist to an issue needing further exploration even if the patient states the topic involved is unimportant or irrelevant’. When a patient is in crisis, fully unpacking these nuances through purely textual inputs—the main form of LLM-chatbot inputs now—is infeasible. Attempting to do so can itself exacerbate distress, particularly if the patient struggles with cognitive or emotional regulation. This should bring home the tension we brought up at the beginning of this section. While fine-tuning promises to bring in the right kind of data, the architecture of LLMs still ends up needing enough data to be able to respond appropriately to all contexts of use. From the foregoing discussion, however, one may start to see the limitations of this way of bridging generality and context-sensitivity: While one can train the model for a variety of contexts, it is not possible to train the model to predict all relevant contexts and adapt to them. For this reason, we advocate for a phased, dynamic process of implementation of LLMs that slowly tests the model in varied contexts of deployment before it is rolled out to more unpredictable ones, as we’ll elaborate in the next section.

It nevertheless remains true that we can significantly ameliorate the problem—we just need more data, but from atypically atypical populations. One drastic solution—though one taken seriously by some—would be to fine-tune the model to the individual case. This is essentially a form of ‘digital phenotyping' \citep{oudin2023digital, tekin2023ethical, bufano2023digital}  involving more elaborate real-time data collection: continuously tracking each patient’s location, biometric signals, social interactions and phone use, to arm the LLM with up-to-date contextual information. Attempts to implement this level of data collection pose significant—and well-known—ethical challenges. First, there are clear risks to patient autonomy and privacy, especially given the tendency of even the best available LLMs to leak private information \cite{mireshghallah2024llmssecrettestingprivacy} and for data leaks to occur; many individuals may consent to occasional check-ins yet baulk at pervasive monitoring of their personal spaces, fearing future data misuse or perpetually living under a clinical microscope. Second, and relatedly, logistical and security challenges loom large: the storage, transmission and interpretation of massive, sensitive data streams would demand robust safeguards—agile and consistent updates to the HIPAA (Health Insurance Portability and Accountability Act) in the United States or the GDPR (General Data Protection Regulation) in the European Union, for example—to guard against leaks, breaches or unethical secondary uses. Third, for certain at-risk patients—especially those with paranoid symptoms—knowing they are under constant observation can itself heighten distress and undermine the therapeutic alliance.

We do not think these rule out the use of LLM-powered chatbots altogether—after all, the person in the snowstorm could not have seen a therapist either. LLM-powered chatbots—suitably curtailed in scope and function—may still offer reprieve. But our discussion is intended to drive home the idea that the atypicality problem does not have easy, one-size-fits-all, solutions: the atypicality problem can appear at all levels in virtue of LLM architecture and can be mitigated but not eliminated. In a sense, every individual encounter is ‘atypical’—and risks the atypicality problem—because each patient’s current episode is shaped by their distinctive environment, life events and symptomatology which may not map neatly onto even the most well-curated datasets.

Of course, not every patient or clinical scenario is so complex, and there may be (many!) contexts where a carefully fine-tuned LLM genuinely enhances psychiatric care (we discuss this in the next section). But especially in high-risk, rapidly evolving scenarios, no amount of fine-tuning can fully resolve the inevitable mismatch between typicality and atypicality—the atypicality problem always looms, and must be an actively managed ethical risk.

\section{Dynamic Contextual Certification}
The problem of atypicality resists technical resolution. Prompt engineering and even individualised fine-tuning cannot eliminate the core architectural limitations that make LLMs ill-suited for decontextualised psychiatric use. If atypicality cannot be solved in the model, it must be managed around the model. To this end, we propose DCC: a staged, iterative and context-sensitive deployment and oversight process for chatbots in psychiatric settings.

DCC draws inspiration from dynamic certification models for autonomous systems \cite{bakirtzis2023}, regulatory strategies for emerging AI technologies \cite{dankslondon2017}, and the staged frameworks used in clinical translation of medical interventions \cite{kimmelman2015structure}. The key idea is that certification of LLMs in psychiatry must not be a one-off test of safety, but a continuous process of context-specific evidence-building and refinement. It assumes (as we tried to show) that systems like LLMs do not simply fail in general, but fail differently across psychiatric subpopulations, depending on evolving local conditions.

DCC is based on what \citet{bakirtzis2023} call the iterative revision of permissible (use, context) pairs. What matters is not whether a system is safe in general, but whether it is safe and effective for a specific population, in a specific context, under specific supervisory conditions. Hence, DCC proceeds in stages, where the scope of the potential system is gradually expanded into broader populations, learning through the process about its limitations and applicability. Ultimately, the objective is to incorporate these systems, but doing so steadily and intelligently.

\begin{enumerate}
    \item \textbf{Phase 1: Pilot Context Evaluation.} In the first stage, the model is deployed in highly controlled, low-risk environments—the most obvious example here would be CBT support for patients with mild depression or anxiety—under close clinical supervision. Drawing from  \citet{kimmelman2015structure}'s work on clinical translation frameworks, the primary task is to locate both where the model is best suited to operate, and where its limits start to show up: where atypical interpretations begin to conflict with therapeutic goals.

    It should be emphasised that this stage is exploratory. Feedback should be gathered from both patients and clinicians, including traceable audit trails,\cite{Falco2021}) such as systematic conversation logging especially with an eye on where adverse situations may have emerged. Crucially, the objective is not simply to demonstrate safety but to gather information that guides the design of later deployments.

    \item \textbf{Phase 2: Targeted Expansion Trials.} Phase 2 expands the system into adjacent psychiatric contexts–for example from patients with chronic depression to patients with comorbid depression and anxiety. The main idea here is to scale up the model gradually and maximise the available data for other conditions. If we have managed a safe deployment for patients with depression, it will be generally more likely that we will successfully scale it up to patients with depression and anxiety than patients with diagnosed schizophrenia. This is based on \citet{dankslondon2017} testing phase for autonomous systems: limited rollouts in real settings, with trained overseers capable of detecting context misalignment.

    An important point here is that transferability of the system is not assumed, but instead being tested in adjacent patient populations. As with drug trials, lateral expansions require prior evidence. A system validated for mild anxiety should not be extended to acute panic disorder or psychosis without specific testing. Where LLMs are concerned, even seemingly adjacent psychiatric conditions can shift how a given output is interpreted. Phase 2, therefore, functions both as a test of generality and as a way to constrain overextension.

    \item \textbf{Phase 3: Conditional General Use.} IIf safety and efficacy are validated in a sufficiently diverse set of contexts, a broader deployment phase may begin—for example, for general clinical use or direct-to-consumer access in predefined psychiatric domains. But this expansion remains conditional: deployment is authorised only where prior (use, context) pairs have been validated. As \citet{Falco2021} argue, black-box flight data recorders work in aviation because failures are auditable in retrospect. The domain of chatbot-assisted psychiatry should be no different, and being able to retrospectively audit the performance of the system is crucial to maintain not only safety, but patient and provider trust.

    At this stage, dynamic certification mechanisms must scale. This is a problem for two reasons. First of all, given the number of stakeholders involved, the certification procedure must include a whole set of processes and institutions (including, for instance, systematic feedback loops with patients and providers, structure audits and traceability logs). Finally, since the interpretability of deep neural networks (particularly profoundly complex systems like transformer architectures) is still limited, it is not clear how models should be evaluated and potentially amended once failure occurs. Since addressing these two problems is not the focus of this paper, we note that this is an open research area worth looking into in the context of dynamic certification of chatbot-assisted services.

    \item \textbf{Phase 4: Continuous Monitoring and Revision.} Most importantly, DCC recognises that certification is never finished. Even if a chatbot proves safe across multiple conditions, many circumstances should lead to re-evaluation, like substantive model updates or new contexts of use. LLMs evolve, patient populations change, and so do their interpretive frames. To put it with \citet{bakirtzis2023}, dynamic certification must ‘learn while certifying’. In the case of DCC, a crucial emphasis is on how new contexts with new interpretative frames must be a crucial part of the feedback loop of learning from deployment.

    New use cases—say, adolescents with mood dysregulation—require their own exploratory phase before being licensed. Similarly, new empirical findings (eg, previously unrecognised risks) may require a rollback of certification, just as pharmaceuticals may lose indications if postmarket trials reveal harm. DCC embeds this reversibility into the structure of deployment itself.
\end{enumerate}
Importantly, DCC does not just gather evidence about system behaviour; it feeds that evidence back into system design and usage policies. This feedback loop is deliberately similar to the relationship between early-phase and late-phase trials in medical research, where exploratory results inform both theoretical understanding and future ensemble design \citep{kimmelman2015structure}. We recognise that such caution will be perceived as an impediment by commercial actors or policy-makers wanting rapid rollout. But the very tension between safety and speed is itself a reason to adopt a phased, reversible approach. Premature deployment can lead to interpretive failures and an irreversible erosion of not just the patient–provider relationship, but trust in the medical institutions more broadly. A dynamic deployment model like DCC is designed to increase the likelihood of long-term acceptance and equitable access–especially for vulnerable populations–by earning trust through steady and intelligent expansion.

One may worry that the population of users with atypical interpretive profiles is small relative to the general population. If so, then the worry is that such a careful rollout may not be justified, and that too much caution might cause more harm than good by delaying access to mental health services. To this, one should start pointing out that approximately 23\% of the American population deals with some form of mental illness (2022 data, \cite{samhsa_nsduh_2023}). WWhile it will not be the case that all of them will have atypical interpretive profiles or require the use of chatbot-assisted therapeutic services, it is worth recalling that the base rate we are dealing with is not with respect to the general population, but relative to the psychiatric subpopulation. Even if considerably below 23\%, the proportion of individuals with some form of mental illness who may face the problem of atypicality is a considerable group of individuals. Furthermore, even if the actual number of individuals affected by the problem of atypicality is small (eg, if only a very small portion is recommended to incorporate chatbots into their treatment), that does not necessarily imply the abandonment of caution. One foundational principle in risk assessment58 is to adjust the strength of precautionary measures in a way that is proportional not just to the probability of negative outcomes, but also their severity: a rare event with severe outcomes—those associated with the problem of atypicality—should nonetheless warrant strong caution.

Additionally, even if the group of actually affected individuals is a small one, their negative experiences can affect the attitudes of the broader population as well, and these influences can persist over time–for instance, marginalised communities in the US continue to mistrust the healthcare system as a result of historical mistreatment. \citep{pascarella_risk_2021} is to adjust the strength of precautionary measures in a way that is proportional not just to the probability of negative outcomes, but also their severity: a rare event with severe outcomes -- those associated with the problem of atypicality -- should nonetheless warrant strong caution. 

Additionally, even if the group of actually affected individuals is a small one, their negative experiences can affect the attitudes of the broader population as well, and these influences can persist over time: for instance, marginalized communities in the US continue to mistrust the healthcare system as a result of historical mistreatment. (\cite{NBERw32028}) The upshot is that trust can be broken even if not all individuals of the psychiatric population are affected by the problem of atypicality, and that once trust is broken, it is very difficult to recover it. In contexts like these, returning to risk assessment, the cost of repairing harm massively surpasses the cost of prevention, which warrants, in our view, a proportionally cautious approach.

\section{Concluding Remarks}


To conclude, the problem of atypicality cannot be solved once and for all. It is not a bug that can be quickly solved, but a structural feature of how LLMs generate language—probabilistically and often without access to the contextual cues that shape psychiatric interaction, especially in the real-time contexts in which chatbots are meant to be used. Our key claim is that we should recognise this fact and adapt our deployment strategy to it. DCC is designed to do just that. It treats atypicality not as a technical hurdle to be resolved once and for all, but a live risk demanding active and contextually sensitive mitigation.

In recent years, there has been a growing push for human-centric or human-centred AI (HCAI), an approach which conceptualises AI systems as tools in service of human needs rather than autonomous replacements \citep{bryson_how_2019, smith_ai_2018, shneiderman_human-centered_2022}.\footnote{We thank an anonymous reviewer for pointing to this connection between our work and HCAI.} This, in part, requires that AI ‘is functioning in a way that can be appropriately controlled and overseen by humans’.\citep{EUAIAct2024} Both \citet{shneiderman_human-centered_2022} and \citet{bryson_how_2019} emphasise the importance of human control through design of appropriate processes and mechanisms, via, for example, governance and regulatory mechanisms. In clinical contexts, we believe that appropriate control must reflect core medical-ethics principles: minimise harm, maximise patient benefit, preserve clinician–patient autonomy and distribute benefits justly.

DCC operationalises these principles. By starting implementation only in narrowly defined domains and contextually constrained, evidence-backed use cases, incorporating active and agile monitoring, and pursuing domain expansion only after safety is demonstrated, DCC lays down one practical path to keeping LLMs under control, in light of the problem of atypicality. Independent oversight and audit trails—features urged by, for example, Bryson’s requirements of accountability, transparency, governance and regulation \citep{bryson_how_2019}, or Shneiderman’s `reliable, safe, trustworthy' framework \citep{shneiderman_human-centered_2022}n HCAI—are built into each rollout phase in DCC. We thus see DCC as a core ingredient for HCAI. Like many working on HCAI, we are not AI sceptics; however, we do think that ‘cooperative coexistence’ \citep{hamid_automation_2017, smith_ai_2018} between AI and humans can only happen when AI implementation is performed responsibly, so that AI can work appropriately for human needs.

\section*{Acknowledgements}
We would like to thank Katie Creel, David Danks, and Frederick Eberhardt for their comments and feedback on previous versions of this paper, as well as Winnie Sung and the audience of the SUTD Symposium on Emerging Intelligence for discussion. An early version was presented at the Te$\chi$nēCon Conference in July 2024, whose audience provided helpful feedback for the paper. We would also like to thank the Ronald and Maxine Linde Center for Science, Society, and Policy for their financial support in the early stages of the project.

\section*{Competing interests}
The authors declare no competing interests.

\bibliography{bib.bib}
\bibliographystyle{unsrtnat}

\end{document}